# Optical left-handed metamaterials made of arrays of upright split-ring pairs

Hsun Chi Chan[1], Shulin Sun[1,2,3,*], Guang-Yu Guo[1,2,*]

[1]Department of Physics and Center for Theoretical Sciences, National Taiwan University, Taipei 10617, Taiwan

[2]Physics Division, National Center for Theoretical Sciences, Hsinchu 3001, Taiwan

[3]Shanghai Engineering Research Center of Ultra-Precision Optical Manufacturing and Green Photonics and Department of Optical Science and Engineering, Fudan University, Shanghai 200433, China

**Electromagnetic metamaterials are man-made structures that have novel properties such as a negative refraction index, not attainable in naturally occurring materials. Although negative index materials (NIMs) in microwave frequencies were demonstrated in 2001, it has remained challenging to design NIMs for optical frequencies especially those with both negative permittivity and negative permeability [known as left-handed metamaterials (LHMs)]. Here, by going beyond the traditional concept of the combination of artificial electronic and magnetic meta-atoms to design NIMs, we propose a novel LHM composed of an array of simple upright split-ring pairs working in the near infrared region. Our electromagnetic simulations reveal the underlying mechanism that the coupling of the two rings can stimulate simultaneously both the electric and magnetic resonances. The proposed structure has a highest refractive index of -2, a highest figure of merit of 21, good air-matched impedance and 180 nm double negative bandwidth, which excel the performances of many previous proposals. We also numerically demonstrate the negative refraction of this metamaterial in both the single-layer form and wedge-shaped lens.**

* Electronic mails: sls@fudan.edu.cn, gyguo@phys.ntu.edu.tw

**I. INTRODUCTION**

An extraordinary substance simultaneously possessing negative permittivity and permeability [dubbed the left-handed material (LHM) or nowadays better known as the double negative metamaterial (DNM)] that satisfies the fundamental physical laws and exhibits novel electromagnetic (EM) properties such as negative refraction, was proposed by Veselago in 1968 [1]. Nevertheless, this fascinating proposal did not receive much attention for a long time because there is no such substance existing in nature with high frequency magnetic response. Since 2001, the research on negative refraction has been reactivated because some artificial metamaterials that possess negative electric and/or magnetic response in controllable frequency regions were fabricated. Metallic wire [2] and split-ring resonator (SRR) [3] as the electric and magnetic meta-atoms are two milestones in this field of EM metamaterials. The first experimental verification of the negative refraction was carried out based simply on a composite structure of these two meta-atoms with overlap EM response frequencies in microwave region [4, 5]. Inspired by this success, various novel EM properties, such as super-lens [6, 7], invisible cloaking [8, 9], and light absorbers [10, 11], have been proposed. The metamaterials demonstrated in the microwave region have been quickly extended to the higher frequencies by scaling down their geometric size [12]. Magnetic response metamaterials have been gradually realized in mid-infrared [13] and near-infrared [14] based on the scaled SRR structure. In optical region, however, metals are no longer perfect conductors and the scaling law thus approaches its limit [15]. The considerable energy dissipation and fabrication challenge of metamaterials have pushed people to look for novel designs. In particular, the optical metamaterial with 'fishnet' structure was experimentally demonstrated to possess a negative refraction with a high figure of merit (FOM, $F = |n'|/n''$) [16]. It consists of the metallic wires and the strip pairs which contribute a negative permittivity and permeability, respectively [17].

Among the diverse proposals of negative index metamaterials, an optical metamaterial of a gold nanorod pair array that was experimentally demonstrated to possess a negative refractive index at the optical communication wavelength of 1.5 μm [18], is particularly important. This interesting proposal breaks the traditional concept of negative index metamaterial composed of electric plus magnetic meta-atoms. Nevertheless, only the real part of permittivity is negative for the gold nanorod metamaterial. Instead, the negative index is realized by a large imaginary part of magnetic permeability via the negative refractive index condition $\varepsilon'|\mu|+\mu'|\varepsilon|<0$ [19]. Such a negative index metamaterial with single negative permittivity has limited frequency band and low FOM [18]. In other words, the 'fishnet' metamaterial can be viewed as an inverse structure of a nanorod pair array and is an archetypal negative index metamaterial in optical region [16]. In addition, the metamaterial of the rod pair plus the continuous metallic wire pair [20] is an alternative with their electric and magnetic response being separately manipulated. These systems go back to the traditional metamaterials made of electric plus magnetic meta-atoms.

SRR is a magnetic meta-atom and the coupling of its split-rings (SRs) offers fascinating possibilities. The first mid-infrared magnetic resonant metamaterial is realized by an array of gold staple-like structure and its mirror image [13]. The laterally coupled SRRs exhibit interesting resonance properties [21-23]. The stereo SRR dimer shows chiral properties [24] and has coupled modes with different orientations [25]. Recently, upright SRR has been successfully fabricated and can be excited by the normal incident EM waves. This fabrication technology offers new opportunities for designing novel three dimensional metamaterials [26].

In this paper, we propose a novel LHD metamaterial made up of only upright split-ring (SR) pairs as meta-atoms. When the material is illuminated by a normal incident wave with the magnetic field perpendicular to the rings, electric and magnetic resonances in the same meta-atom will be excited simultaneously in the optical frequency region. Numerical EM simulations reveal a negative index of -1.9 around the wavelength of 1200 nm with air-matched impedance in our system. The highest FOM is about 21 and the band width of the negative index is about 250 nm. Compared with previous proposals [16-20], such performance is excellent, and also counter-intuitive considering that a double negative metamaterial can be realized by the well-known magnetic meta-atoms.

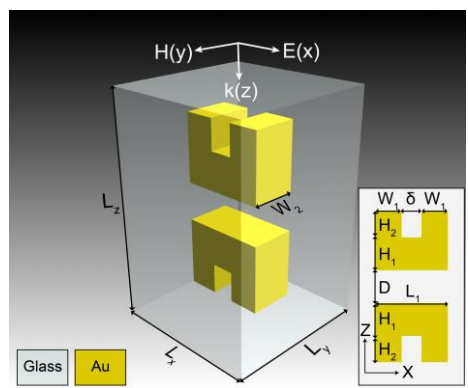

**FIG. 1. Proposed LHM structure.** The unit cell consists of two back-to-back upright gold (yellow) spit-rings surrounded by glass. The lattice constants of the DNM are 200 nm ($L_x$) × 200nm ($L_y$) × 300nm ($L_z$). Other parameters are: $W_1 = 40 nm, W_2 = 60 nm, H_1 = 50 nm, H_2 = 40 nm, D = 50 nm, \delta = 30 nm$. The EM wave is normally incident on the structure with the electric (magnetic) field along $x$ ($y$) direction.

## II. RESULTS AND DISCUSSION

Our proposed LHM consists of a square array of upright SR pairs embedded in a slab of homogeneous glass, as illustrated in Fig. 1. The two rings in each SR pair stand back-to-back with a gap $D = 50$ nm and can be excited by a normal incident EM wave with its electric (magnetic) field parallel (perpendicular) to the ring plane. The small separation of the two SRs introduces a significant coupling effect, and this plays an important role to realization of the LHMs. The lattice constants of the proposed structure are 200 nm ($L_x$) × 200 nm ($L_y$) with slab thickness of 300 nm ($L_z$). It is not so sub-wavelength in the wave propagation $z$-direction because two SRs lined up along this direction with inserted dielectric layers (the working wavelength is around 1200 nm). An advantage is the relatively low volume ratio (8.7 %) of metal in the unit cell which may lead to a low-loss metamaterial. Such a design could be fabricated by electron beam lithography [26] and be further simplified with its half region replaced by the mirror image [13].

We perform numerical EM simulations based on the finite element method implemented in COMSOL Multiphysics® software to study the EM properties of the proposed LHM. The permittivity of gold is described by the Drude model $\varepsilon_{Au} = 9 - \omega_p^2 / [\omega(\omega + i\Gamma)]$ with $\omega_p = 1.37 \times 10^{16}$ Hz and $\Gamma = 1.0027 \times 10^{14}$ Hz [18]. The refractive index of glass is 1.458. Periodic boundary conditions are applied in the *x*- and *y*-direction (Fig. 1) to simulate a single layer LHM with infinite lateral dimensions. Figure 2(a) shows the calculated transmittance (*T*) and transmission phase shift ($\Delta\phi$) of the proposed LHM as a function of wavelength. The $\Delta\phi (= \phi_s - \phi_r)$ refers to the difference of the transmission phases introduced by a sample ($\phi_s$) and a reference layer of air with the same thickness ($\phi_r$) when the EM waves transmit through them [18]. Furthermore, the real parts of the permittivity $\varepsilon_{eff}$ and permeability $\mu_{eff}$ are retrieved from the calculated amplitude and phase of both the transmission and reflection, as displayed in Fig. 2(b) [27]. It should be pointed out that while single SR is bianisotropic [27], our SR pair system possesses no bianisotropy since it is symmetric in the propagating direction, provided that no spatial dispersion is assumed. Therefore, the effective parameters obtained from the bianisotropic retrieval method [28] and the original isotropic one [27] would make no difference.

Figure 2(a) shows that the transmittance dips at around 1008 nm, where the phase shift $\Delta\phi$ also shows a resonant behavior. This indicates that the phase accumulations of the EM waves inside the metamaterial and the reference (air) have opposite signs, which is a footprint of the negative refraction [18]. Furthermore, in the wavelength region between 1080 and 1260 nm, both $\text{Re}(\varepsilon_{eff})$ and $\text{Re}(\mu_{eff})$ are negative, as displayed in Fig. 2(b), and thus the proposed structure is a LHM. It is well known that the impedance mismatch is a common problem because the electric resonance of the negative refraction metamaterials is usually stronger than the magnetic one. Here the proposed LHM possesses a good tradeoff between the electric and magnetic resonances which has similar bandwidth and amplitude. Since the glass occupies 91% of the unit cell volume and supplies a positive background permittivity, this design will suppress the effective negative permittivity of the whole LHM.

We also calculate the effective index $n_{eff}$ and impedance $Z_{eff}$, with their real parts shown in Fig. 2(c). We can see that the system behaves as an effective negative refraction metamaterial [NRM, see Fig. 2(a)] in the wavelength region between 1008 and 1260 nm, which can be further categorized to the LHM region between 1080 and 1260 nm and the single negative metamaterial (SNM) region between 1008 nm and 1080 nm. The negative index of the SNM is also contributed by a large imaginary part of magnetic permeability $\mu_{eff}$ which leads to a large energy loss and low FOM [19]. Such analysis is supported by the calculated FOM spectrum [Fig. 2(d)] which is smaller than 1 in the SNM region. Furthermore, the impedance of SNM is considerably mismatched with air and the transmittance is therefore small (see Fig. 2). For the LHM region, the real refractive index is smaller than -1 in the whole region and the FOM is much larger. Around 1204 nm, the refractive index is about -2.0 and the FOM is about 20.6. This is consistent with the above analysis that the low metal volume ratio (7.8 %) in our structure will contribute to realize a low-loss metamaterial. On the other hand, since the 1204 nm wavelength is far away from the electric and magnetic resonance wavelengths (see Fig. 3), the Ohmic loss gets further reduced, also contributing to the high FOM value. Besides, the impedance of the LHM matches quite well with that of air. Therefore, these calculated properties are advantageous Therefore, these calculated properties are advantageous relative to the referenced structures [16-20].

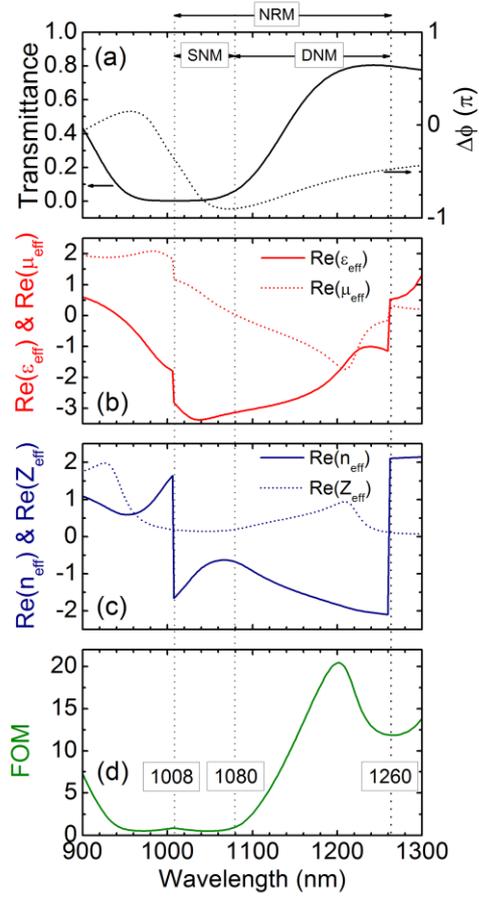

**FIG. 2. Calculated spectra of transmittance and effective optical parameters of the proposed LHM.** (a) transmittance ($T$) / transmission phase shift $\Delta\phi$, (b) the real parts of the retrieved effective permittivity ($\varepsilon_{eff}$) / permeability ($\mu_{eff}$), (c) effective refractive index ($n_{eff}$) / impedance ($Z_{eff}$) and (d) the FOM. Here NRM, SNM and DNM denote negative refraction metamaterial, single negative metameterial and double negative metamaterial, respectively. The discontinuities in the Re ($\varepsilon_{eff}$), Re ($\mu_{eff}$) and Re ($n_{eff}$) spectra in (b) and (c) are explained in Appendix A.

Electric- and magnetic-responses are tailored to overlap for the realization of DNM. The proposed structure of the gold SR pairs can serve as both the electric and magnetic meta-atoms stemming from their coupling effect. To see this, we show in Fig. 3(a) the transmittance spectra of the same structure as in Fig. 2(a) with different arm lengths $H_2$ of the upright SR. At $H_2 = 0$ nm, two transmittance dips appear at 705 and 827 nm, corresponding to two coupled resonance modes. As $H_2$ increases, the two dips will red shift and move closer to each other. Such tendency can be clearly seen in Fig. 3(b) where the wavelengths of the two transmittance dips versus the arm length of the SR pair $H_2$ are plotted. The red shift of the two transmission dips can be attributed to the fact that the flowing paths of the induced currents inside the rings become longer as $H_2$ increases, because the EM resonance wavelength is proportional to the distance of such loops. Our numerical simulations indicate that these two dips correspond to resonant mode 1 (long-wavelength) and mode 2 (short-wavelength) with anti-parallel and parallel induced currents inside the two SRs [see the insets in Fig. 3(b)]. For mode 1 (2), the magnetic (electric) responses in the two rings are parallel with its electric (magnetic) response almost cancelled. With increasing $H_2$, these two resonances overlap at about $H_2 = 40$ nm and the condition of the negative refraction metamaterial is satisfied. The electric- and magnetic-responses of the proposed system are anti-parallel to the electric and magnetic fields of the incident EM waves, and the double negative refraction is thus realized.

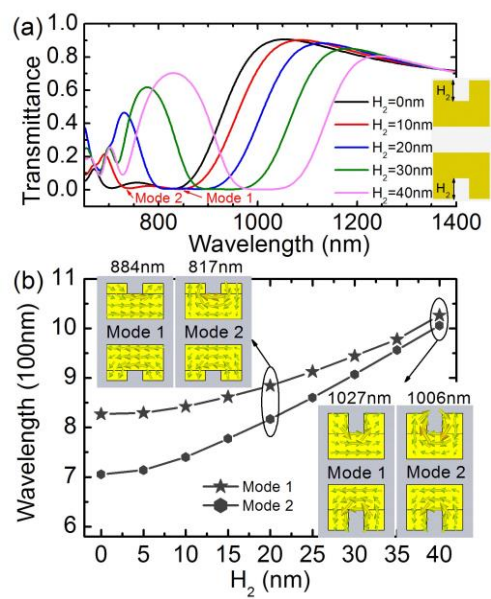

**FIG. 3. Calculated transmission spectra and resonance wavelengths of the proposed DNM.** (a) Transmittance spectra with different arm lengths of the SRs $H_2$.

The wavelengths of the transmission dips are related to two different modes labelled by mode 1 and mode 2. (b) Resonance wavelengths versus the arm length $H_2$ for mode 1 (star line) and mode 2 (hexagon line). The induced electric current distributions inside the two SRs are shown as inset in (a) for two modes in two systems with different $H_2$.

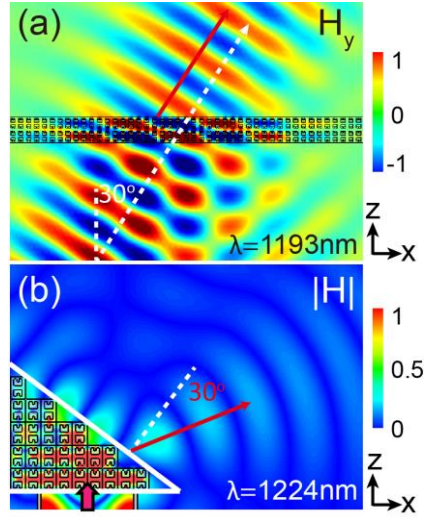

**FIG. 4. Demonstration of the negative refraction property of the proposed metamaterial.** (a) Calculated $H_y$ field distribution of a $30°$ incident TM polarized Gaussian beam transmitting through a two-layer DNM. The central lines of the incident beam and transmitted beam are marked by the dark and red lines, respectively. (b) Calculated magnetic field $|\vec{H}|$ of the TM polarized plane wave transmitting from a waveguide through a wedge-shaped lens made of the proposed DNM. The refracted beam (red line) stays at the same side of the normal line (white dashed line) with the incident beam (pink arrow). The field patterns of (a) and (b) are chosen at the central cutting plane of the double SRs and the perfect magnetic conductor (PMC) boundary conditions are set along $y$ direction. The background medium of both cases is air.

To further understand the coupling mechanism, here we employ a Lagrangian model, regarding the SR pair as two LC circuits with resonant frequency $\omega_0 = 1/\sqrt{LC}$ and neglecting the Ohmic loss for simplicity. The Lagrangian for the SR pair can be written as [25] $\mathcal{L} = L(\dot{Q}_1^2 - \omega_0^2 Q_1^2)/2 + L(\dot{Q}_2^2 - \omega_0^2 Q_2^2)/2 + M_M \dot{Q}_1 \dot{Q}_2 - M_E \omega_0^2 Q_1 Q_2$, where generalized coordinates, $Q_{1,2}(t)$, are the charges accumulated in the gap of the two SRs. $M_m$ and $M_e$ are the mutual inductance and mutual capacitance, respectively. By substituting the Lagrangian $\mathcal{L}$ into the Euler-Lagrange equation, we obtain two eigenfrequencies for symmetric and antisymmetric modes as $\omega_S = \omega_0 \sqrt{\frac{1+\kappa_E}{1+\kappa_M}}$ with $Q_1 = Q_2$, and $\omega_{AS} = \omega_0 \sqrt{\frac{1-\kappa_E}{1-\kappa_M}}$ with $Q_1 = -Q_2$, where $\kappa_E = M_E/L$ and $\kappa_H = M_H/L$ are dimensionless electric and magnetic coupling constants, respectively. As can be seen in Figure 3(a), these results explain the splitting of the resonant frequency, i.e., we can assign $\omega_{AS}$ ($\omega_S$) to mode 1 (mode 2) with the corresponding anti-parallel (parallel) induced current. In particular, the Lagrangian model shows clearly that $\kappa_E$ and $\kappa_H$ would approach to each other as $H_2$ increases, and eventually become equal (i.e., no splitting and $\omega_{AS} \simeq \omega_S \simeq \omega_0$) at $H_2 = 40$ nm.

To demonstrate that the negative refraction phenomenon indeed occurs in the proposed DNM, we further perform three dimensional numerical simulations. Figure 4(a) shows that a transverse magnetic (TM) polarized Gaussian beam $\vec{H}(x,z) = e^{-r^2/w^2} e^{i(k_0 z \cos\theta - k_0 x \sin\theta)} \hat{y}$ transmits through a double-layer DNM from air. Here $w = 2$ μm is the beam width, $r$ is the radial distance to the central axis of the beam, the incident angle $\theta$ is $30°$ and the wavelength $\lambda$ ($= 2\pi/k_0$) is 1193 nm. Clearly, the transmitted beam is shifted backwards (along -$x$ direction) because it encounters a negative refraction inside the DNM. The reflected beam is almost negligible because the DNM has a good impedance matching with air [see Fig. 2(c)]. Although the subwavelength property of the proposed DNM is not perfect ($\sim \lambda/4$), one major advantage is that the negative refraction effect is visible even in the structure composed of two or few layers. Furthermore, numerical experiments are also carried out for a wedge-shaped lens [marked by the white lines in Fig. 4(b)] made of an array of gold SR pairs. As shown in Fig. 4(b), the refracted and incident beams appear at the same side of the normal line of the wedge-shaped lens. Here the TM polarized incident beam is transported by a waveguide and shined on the bottom of the lens. The negative refraction phenomenon can be observed in a broad wavelength region and the wavelength of 1224 nm is chosen in Fig. 4(b). According to the measured refracted angle (about $30°$) and the incident angle (equals to the inclined angle of the lens, $38.7°$), the effective refractive index of the lens device is about -0.8, which is smaller than the retrieved value of the single layer DNM (about -2). Because of the inevitable interaction between the different layers inside the bulk metamaterial, the retrieved effective medium properties based on a single layer case is thus different from the bulk system. The low transmitted intensity can be attributed to the inter-layer coupling, increased metal loss and impedance mismatch. Importantly, the negative refraction effect survives in the composite device. And we have also numerically demonstrated that this layer-layer interaction can be suppressed by, e.g., introducing the displacement between the different layers.

Usually, to prove a proposed metamaterial that possesses a negative refraction index, the dispersion relations of the structure are calculated as an important piece of evidence [16]. Similarly, we calculated the dispersion relations of the infinite metamaterials made by either the single ring or upright SR pair, as shown in Figs. 5(a) and (b), respectively. For the single SR case, a band gap exists in the band of about 230 - 300 THz (i.e., 1000 – 1300 nm). Interestingly, for the SR pair case, a transparent photonic band with negative group velocity (around 1120 – 1210 nm) clearly appears inside the band gap region of the single SR system, indicating the negative refraction property of the SR pair metamaterial. Here, the negative index band is narrower than that shown in Fig. 2. This is due to the fact that the dispersion relations were calculated based on a bulk metamaterial while the effective medium parameters shown in Fig. 2 were based on a single unit cell only. To validate this argument, we also calculated the transmission spectrum of one, three and seven split-ring pairs.

Figure 5(c) shows that as the number of unit cells increases, the transmission spectrum matches better with both the propagation and stop bands because the bulk properties of the metamaterials are approached. Moreover, we calculated the phase distribution along the propagation direction inside a four-layer metamaterial surrounded by the air at two specific wavelengths of 1176 nm and 908 nm [29], demonstrating clearly a negative and positive phase velocity inside the metamaterials (see Fig. 6 in the Appendix).

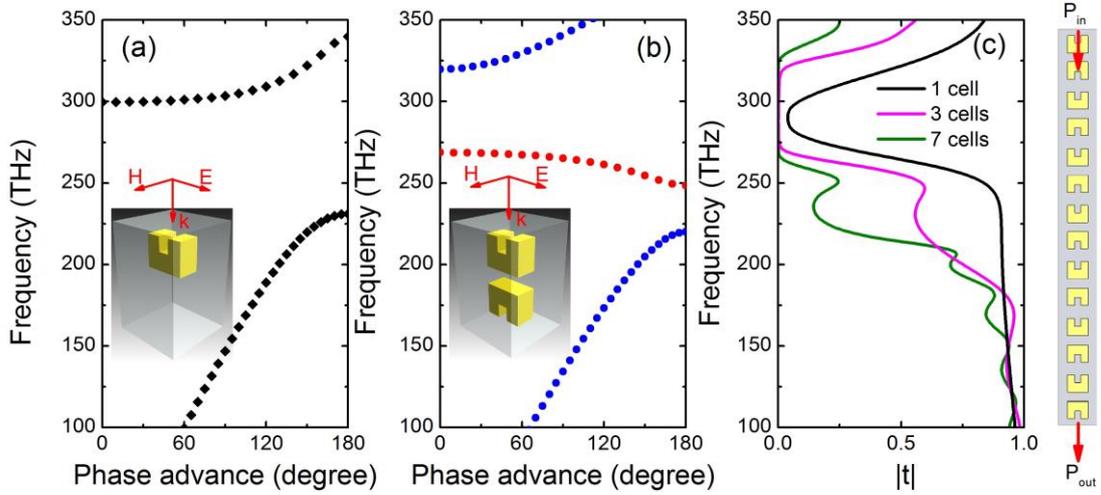

**FIG. 5. Calculated dispersion relations and transmission of the infinite metamaterials.** (a) Single ring array and (b) upright split-ring pair array. (c) Transmission spectrum of the upright split-ring pair metamaterial with one, three and seven unit cells. In the calculations, periodic boundary conditions are set along the electric and magnetic field directions, and Floquet boundary condition is set along the wave vector direction with the swept wave vector $k_z$.

## III. CONCLUSIONS

In summary, we have proposed a novel LHM based on gold SR pairs that would have a double negative index in a broad band between 1080 and 1260 nm. The system would have a largest effective refractive index of about -2, a highest FOM of about 21 and an air-matched impedance. In contrast to the traditional concept that a negative index metamaterial consists of electric plus magnetic meta-atoms, the proposed meta-atom has overlap electric and magnetic resonance modes due to the coupling effect between the two SRs. Our numerical simulations have further demonstrated the negative refraction effect in both a single-layered metamaterial and a wedge-shaped lens device made of the gold SR pairs. In addition, the calculated dispersion relations also confirm the negative refraction property of the proposed metamaterial. Our findings thus provide a novel route to design LHMs and also NIMs, thus contributing to the advances in the growing field of three dimensional optical metamaterials.

## Acknowledgments

This work is supported by the Ministry of Science and Technology and the National Center for Theoretical Sciences of Taiwan as well as the Natural Science Foundation of China (No. 11404063) and Shanghai Science and Technology Committee (Grant no. 14PJ1401200), Shanghai, China.

## Appendix A: Comments on the Retrieval Method

It should be noticed that the discontinuities in the retrieved spectra of the effective optical parameters, such as those shown in Figs. 2(b) and 2(c), is an intrinsic problem of the S-parameter retrieval method [27]. This is because the effective optical parameters are a multi-valued function of physical quantities [e.g., reflection ($r$) and transmission ($t$) amplitudes as well as slab thickness ($d$)] and thus have many branches. For example, the effective refraction index is given by

$n = \pm \cos^{-1}\left[(1-r^2+t^2)/(2t)\right]/(k_0 d) + 2m\pi/(k_0 d)$, where $m$ = 0, ±1, ±2, etc. If a specific branch number is chosen for the whole wavelength range, the discontinuities would occur [see, e.g., Figs. 2(b) and 2(c)]. Moreover, there is no unique way for the selection of branch number and thus one has to select a meaningful branch number based on physical considerations.

In this work, branch number $m = 0$ is chosen for the whole considered wavelength regime because the single layer in our metamaterials is of subwavelength thickness and therefore the transmission phase of the EM waves would not exceed $2\pi$.

This, however, leads to the jumps in the retrieved effective optical parameters at several different wavelengths [see Figs. 2(b) and 2(c)], as expected. Nevertheless, this strategy of choosing the branch number has been widely adopted (see, e.g., [17,30-31]). Moreover, physical effects such as the resonant phase shift of the single-layer metamaterial (Fig. 2a), the light bending effect in a two-layer slab and wedge-like structures (Fig. 4), the calculated dispersion relations (Fig. 5) and the averaged phase evolution (Fig. 6) presented in the main text and Appendix B lend further support to this choice of branch number.

## Appendix B: Supplementary Figures

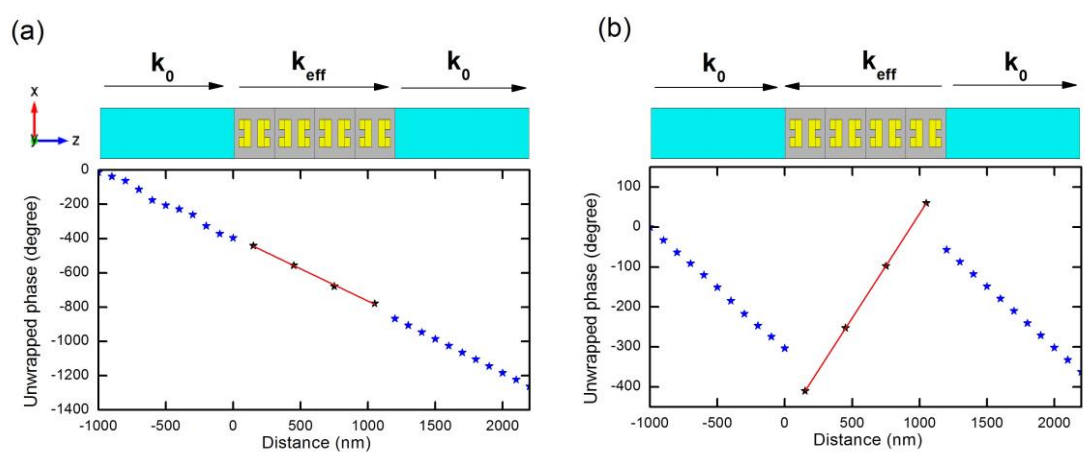

**Figure 6. Calculated EM wave phase as a function of distance along the propagation direction inside the four-layer metamaterial in air by using the wave propagation retrieval method.** (a) At the 908 nm wavelength and (b) at the 1176 nm wavelength. Note that the phases inside the inhomogeneous metamaterial regions (black dots) are obtained by integrating the averaged $E_x$ field over a unit cell [29]. The real part of effective index of refraction, given by the slope of the phase-distance curves (the red fitted lines), is (a) $\text{Re}[n_{\text{eff}}] = 0.95$ and (b) $\text{Re}[n_{\text{eff}}] = -1.7$, agreeing well with those obtained by the usual retrieval method [27] for our four unit cells system.